\documentclass[preprint,prl,showpacs,preprintnumbers,amsmath,amssymb,floatfix]{revtex4}
\usepackage{amsmath}
\usepackage{lipsum}
\usepackage{graphicx}
\usepackage{dcolumn}
\usepackage{bm}
\usepackage{ulem} 
\usepackage[usenames]{color}
\usepackage{epstopdf}
\usepackage{epsfig}
\usepackage{float}
\usepackage{subfigure}
\usepackage{multirow}
\usepackage[version=3]{mhchem}

\usepackage{soul}

\newcommand{\nc}{\newcommand}       
\nc{\vc}[1] {\mbox{\boldmath $#1$}} 
\nc{\del}       {\partial}              
\nc{\bra}       {\langle}               
\nc{\ket}       {\rangle}               
\nc{\bras}[1]   {\langle #1|}           
\nc{\kets}[1]   {|#1\rangle}            
\nc{\mapleft}[1]{           
 \smash{\mathop{\,          %
  \hbox to 1.5cm{\rightarrowfill}\, }\limits_{#1}}}
\nc{\beq}     {\begin{eqnarray}} \nc{\eeq}    {\end{eqnarray}}
\nc{\nn}      {\\\nonumber} \nc{\vs}      {\vspace{-0.275cm}}
\nc{\fra}    {\frac{1}{2}}
\nc{\mb}        {\mathbf}

\begin{document}

\preprint{}

\title{Resolving the spurious-state problem in the Dirac equation with finite difference method}

\author{Ying Zhang}
\affiliation{Department of Physics, School of Science, Tianjin University, Tianjin 300354, China}

\author{Yuxuan Bao}
\affiliation{School of Physics, Nankai University, Tianjin 300071,  China}

\author{Jinniu Hu}~\email{hujinniu@nankai.edu.cn}
\affiliation{School of Physics, Nankai University, Tianjin 300071,  China}
\affiliation{Shenzhen Research Institute of Nankai University, Shenzhen 518083, China}

\author{Hong Shen}~\email{songtc@nankai.edu.cn}
\affiliation{School of Physics, Nankai University, Tianjin 300071,  China}

\date{\today}
\begin{abstract}
To solve the Dirac equation with the finite difference method, one has to face up to the spurious-state problem due to the fermion doubling problem when using the conventional central difference formula to calculate the first-order derivative on the equal interval lattices.  This problem is resolved by replacing the central difference formula with the asymmetric difference formula, i.e., the backward or forward difference formula.  To guarantee the hermitian of the Hamiltonian matrix, the backward and forward difference formula should be used alternatively according to the parity of the wavefunction.  This provides a simple and efficient numerical prescription to solve various relativistic problems in the microscopic world.  
\end{abstract}



\maketitle
The Dirac equation is essential to describe the relativistic systems consisting of spin one-half particles in atomic physics, nuclear physics, and particle physics.  The Dirac equation is a coupled first-order differential equation for the wavefunctions with large and small components.  It can be solved  analytically with very few potentials. Therefore, the numerical methods to obtain the eigen energies and wavefunctions of the Dirac eqation are highly demanded in the relevant fields. Many numerical technologies have been applied to solve the Dirac equation, such as shooting method~\cite{horowitz81,meng06}, basis expansion method~\cite{gambhir90,zhou03,geng07,lu14}, finite difference method (FDM)~\cite{salomonson89,zhao16,fang20}, finite element method (FEM)~\cite{bottcher87,fischer93,poschl96,muller98,zhao07,almanasreh13}, Green's function method~\cite{sun20}, imaginary time step (ITS) method~\cite{ying10}, inverse Hamiltonian method~\cite{hajino10,tanimura15}, conjugate gradient method~\cite{li20}, and so on. {{Among these methods, the shooting and basis expansion methods are quite robust.  They are extensively used to solve the Dirac equation in the relativistic mean-field model, which is a powerful tool to describe many nuclear properties~\cite{meng06,ring96,vretenar05,meng15,meng16}}.  However, these two methods are very sensitive to the box size or basis space for the weakly bound states}. The Green's function method is not sensitive to the space size but cannot give the eigen energies and wavefunctions directly.  The ITS, inverse Hamiltonian and conjugate gradient methods are also friendly to the space size but they need steps of evolutions to achieve the final solutions.  

The FDM is a very simple and efficient method to solve the differential equation, where the derivative operator is easily replaced by a combination of several function values with the finite difference formula.  This method does not need any evolution process.  It achieves great success in solving the {Schr\"{o}dinger} equation~\cite{bartlett52}. In the lattice quantum chromodynamics (LQCD) theory~\cite{susskin77,stacey82}, people found a so-called `fermion doubling' problem when the Dirac field is discretized with a central difference formula (CDF), i.e., more fermionic states than expected were obtained.  On the other hand, one could get the spurious solutions with rapidly oscillating wavefunctions mixing up with the physical solutions in solving the Dirac equation with FDM due to the same reason.

In LQCD, it is tried to remove the fermion doubling problem by introducing an external energy term in the Hamiltonian, i.e., the Wilson term, which modifies the energy-momentum dispersion relation of Dirac particle and shifts the spurious state to the continuum spectrum~\cite{wilson77,kogut1983}. Alternatively, the high-accurate finite difference formula for the first-order derivative with more lattice points can also help to reduce the number of spurious states in Dirac equation~\cite{salomonson89,fang20}.  

In this work, we will propose a novel and simple prescription to solve the spurious-state problem when solving the Dirac equation with the FDM for massive fermions without adding any artificial terms.  For a numerical illustration, we will take the nucleons of a finite nucleus moving in a Dirac Woods-Saxon potential as an example to explain this prescription.   

The Dirac equation describing a nucleon with the mass $M$ moving in the scalar $S(\bm{r})$ and vector $V(\bm{r})$ potentials can be written as~\cite{meng06},
\beq
\{\bm{\alpha}\cdot\bm{p}+V(\bm{r})+\bm{\beta}[M+S(\bm{r})]\}\Psi(\bm r)=\varepsilon \Psi(\bm r),
\eeq
where, $\bm\alpha$ and $\bm\beta$ are the Dirac matrices, $\varepsilon$ and $\Psi(\bm r)$ are the eigen energy and the corresponding wavefunction, respectively.

In a spherical system, the wavefunction can be written as,
\beq
\Psi(\bm r)=\frac{1}{r}
\begin{gathered}
	\begin{pmatrix}
		G(r)Y_{ljm}\\iF(r)Y_{\tilde{l}jm}
		\end{pmatrix}
\end{gathered},
\eeq
where $l=j\pm1/2$ and $\tilde{l}=2j-l$. $G(r)$ and $F(r)$ are the large and small components of the wavefunction, respectively. $Y_{ljm}(\hat{\bm r})$ is the spin spherical harmonics.
Therefore, the radial Dirac equation can be obtained as,
\beq\label{rdir}
\begin{gathered}
	\begin{pmatrix}
		\Sigma(r)&\frac{\kappa}{r}-\frac{d}{dr}\\\frac{\kappa}{r}+\frac{d}{dr} & \Delta(r)
	\end{pmatrix}
\begin{pmatrix}
	G(r)\\F(r)
\end{pmatrix}
=E\begin{pmatrix}
	G(r)\\F(r)
\end{pmatrix}
\end{gathered}, \label{eq:Dirac-matrix-1}
\eeq
where 
\beq
&&\Sigma(r)=V(r)+S(r),   \nn
&&\Delta(r)=V(r)-S(r)-2M,  \nn
&&E=\varepsilon-M,  \nn
&& \kappa=(-1)^{j+l+1/2}(j+1/2).
\eeq
We take the Woods-Saxon potentials describing the finite nuclei for the $\Sigma(r)$ and $\Delta(r)$ fields from  the relativistic mean-field model. The details can be found in Ref.~\cite{koepf91}. Explicitly, we take the neutron in the nucleus $^{132}{\rm Sn}~~(N=82,~Z=50)$ as an example in the following calculation.   

With the FDM, the first-order derivative operator $d/dr$ in Eq.~(\ref{rdir}) can be replaced by a numerical differentiation formula on the equal interval lattices.  Then the Dirac Hamiltonian in Eq.~(\ref{rdir}) can be expressed as a matrix in the coordinate space.  The eigen energies and wavefunctions can be easily obtained by diagonalizing this Dirac Hamiltonian matrix. 

There are many formulas for the finite difference approximations to calculate the first-order derivative. 
The 3-point CDF is the simplest one that approximates the first-order derivative of a function $f(r)$ at $r$ by,
\beq\label{CDFF}
\frac{df(r)}{dr}\simeq\frac{f(r+h)-f(r-h)}{2h},
\eeq
where, $h$ is the lattice interval.   If the position $r$ is equally discretized as $n$ lattices,  the first-order derivative of $f(r)$ can be written in a matrix form,
\beq
\frac{d}{dr}=\frac{1}{2h}
\begin{gathered}
	\begin{pmatrix}
		0&1& & & & & & \\
	  -1&0&1 & & & && \\
	     & & &... & & & &\\
	      & & & &  &-1 &0 & 1\\
	      & & & &  &   &-1& 0\\
	\end{pmatrix}.
\end{gathered} \label{eq:dfdr-3p-CDF}
\eeq
In the following calculation, we take a box with $R_{\rm box}=20$~fm and $n=500$ lattices. 
{Furthermore, we assume the boundary condition for the wavefunctions as 
$f(r)=0$, for $r=0$ and outside the box, $r>R_{\rm box}$}.
Then, the Dirac equation (\ref{eq:Dirac-matrix-1}) can be written in the matrix form as
\beq
\left(\begin{array}{c|c}
	\bm{A}&\bm{B_1}\\
	\hline
	\bm{B_2}&\bm{C}
	\end{array}\right)
\left(\begin{array}{c}
	\bm{G}\\
	\bm{F}
\end{array}\right)
=E\left(\begin{array}{c}
	\bm{G}\\
	\bm{F}
\end{array}\right), \label{eq:Dirac-matrix-2}
\eeq
where $\bm G$ is a vector for the large component of wavefunction, $G(r)$ at $r_1=h,~r_2=2h,~...,~r_{n-1}=(n-1)h,~r_n=nh$,
\beq
\bm{G}=\left(\begin{array}{c}
	G(r_1)\\
	G(r_2)\\
	...\\
	G(r_{n-1})\\
	G(r_{n})
\end{array}\right)
\eeq
and $\bm F$ has a similar structure as $\bm G$ but for the small component. The matrices $\bm A$ and $\bm C$ are diagonal with $\Sigma(r)$ and $\Delta(r)$.
%
The matrix $\bm B_1$ can be written as
\beq
\bm B_{1c}=
\begin{gathered}
	\begin{pmatrix}
		 \frac{\kappa}{r_1}&-\frac{1}{2h}& & & & & & \\
			\frac{1}{2h}&\frac{\kappa}{r_2}&-\frac{1}{2h}& & & && \\
		& & &... & & & &\\
		& & & &  &  \frac{1}{2h}&\frac{\kappa}{r_{n-1}}&-\frac{1}{2h}\\
		& & & &  &   & \frac{1}{2h}&\frac{\kappa}{r_{n}}\\
	\end{pmatrix}
\end{gathered}
\eeq
and $\bm B_{2c}=\bm B^T_{1c}$.  

\begin{figure}[htb]
	\centering
	\includegraphics[scale=0.38]{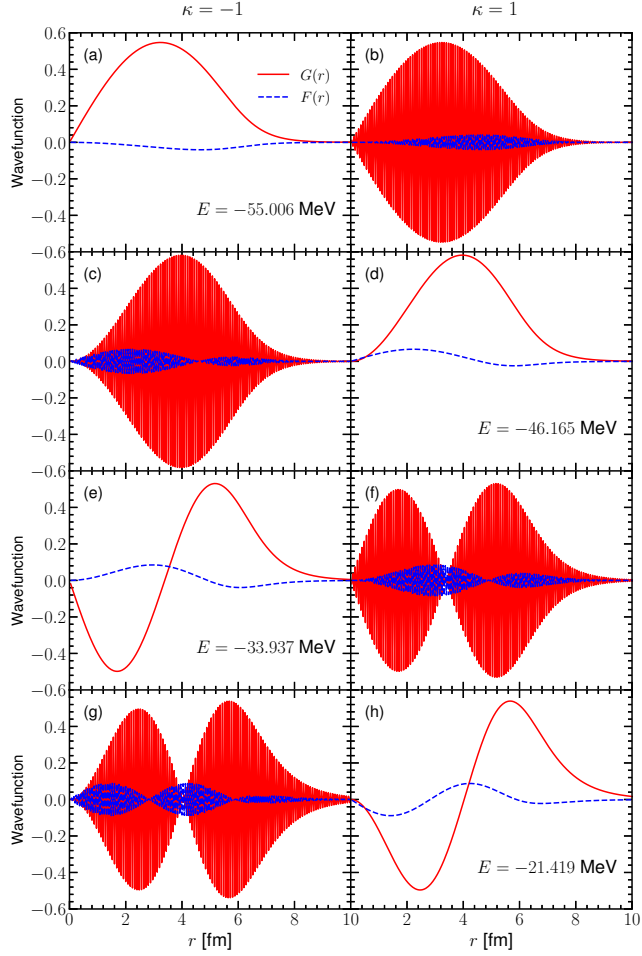}
	\caption{The wavefunctions of neutrons in $^{132}{\rm Sn}$ for the states with $\kappa=-1$ and $\kappa=1$ in Wood-Saxon potential obtained by the FDM with 3-point CDF.}\label{fig.1}
\end{figure}

After diagonalizing the above Dirac Hamiltonian matrix (\ref{eq:Dirac-matrix-2}), one can get $n$ sets of eigen energies and the corresponding wavefunctions.  
The first five bound states obtained by the above FDM using the 3-point CDF for the states $ns_{1/2}$ ($\kappa=-1$) and $np_{1/2}$ ($\kappa=1$) are shown in the columns 3PCDF in Table~\ref{table.1}.   For comparison, the results obtained by the shooting method are also listed in the same table.  One can find pairs of degenerate solutions between $\kappa=1$ and $\kappa=-1$.  The large and small components of their wavefunctions are shown in Fig.~\ref{fig.1}.  The panels (a) and (b) of Fig.~\ref{fig.1} show the wavefunctions of the states with the same energy $E=-55.006$~MeV obtained for $\kappa=-1$ and $\kappa=1$, respectively.  It is easy to identify the physical state $1s_{1/2}$ in panel (a) for $\kappa=-1$, but the spurious state with rapidly oscillating wavefunctions is observed in panel (b) for $\kappa=1$.  
Similar spurious states appear in panels (c), (f), and (g) for the states  with energies $E=-46.165$~MeV ($\kappa=-1$), $E=-33.937$ MeV ($\kappa=1$), and $E=-21.419$ MeV ($\kappa=-1$), respectively.  All the spurious states are marked in boxes in Table~\ref{table.1}.  On the other hand, the physical solutions have the close energies to those obtained by the shooting method.  

Actually, the origin of the above degenerate physical and spurious states has been demonstrated in Ref.~\cite{zhao16}.   {He} found if the first-order derivative is calculated by the 3-point CDF as in Eq.~(\ref{eq:dfdr-3p-CDF}), there exists a unitary matrix $U$ that transforms the Hamiltonian with $\kappa$, $H_\kappa$ to that with $-\kappa$, $H_{-\kappa}$, i.e., $UH_\kappa U^{-1}=H_{-\kappa}$.   This matrix $U$ has alternative $\pm1$ diagonal elements.   As a result, one can obtain the degenerate energy solutions $E_{\kappa}=E_{-\kappa}$, with the wavefunctions $\phi_{\kappa}=U\phi_{-\kappa}$.  If the wavefunction $\phi_{-\kappa}$ is a physical solution, the corresponding $\phi_{\kappa}$ will have rapidly oscillating wavefunction between the positive and negative values, and thus becomes a spurious solution.  This can be seen in the panels (b), (c), (f), and (g) of Fig.~\ref{fig.1}.  However, the half of each envelops of these oscillating wavefunctions are identical to those of the physical state with the same energy. 

One may try to use the 5-point CDF that has a higher accuracy to calculate the first-order derivative instead. The obtained results for $\kappa=-1$ and $\kappa=1$ are listed in columns 5PCDF in Table~\ref{table.1}.  In this case, one can find that the degeneracy between the physical and spurious solutions disappear.  This is because there is no such unitary matrix $U$ to transform $H_{-\kappa}$ to $H_{\kappa}$ anymore.  Therefore, the number of spurious states is reduced compared to those obtained by the 3PCDF.  This fact was also found in Ref.~\cite{salomonson89}.  

To avoid the fermion doubling problem, Ref.~\cite{stacey82} used the 2-point forward or backward difference formula, i.e. asymmetric difference formula (ADF) to discretize the Dirac field of the massless fermion in one-dimensional  LQCD.  Recently, Ref.~\cite{szafran19} clearly pointed out in their Figure 5 that the central symmetric formula (\ref{CDFF}) uses the wavefunctions at $r-h$ and $r+h$ to calculate the first-order derivative, but misses the information at $r$.  This can also explain the fact that 5PCDF can produce less spurious states since it misses less information at $r$ comparatively.   Therefore, Ref.~\cite{szafran19} applied the ADF to calculate the first-order derivative in the mesh-sweeping method to solve the Dirac equation for electrons in two-dimensional graphene.  

\begin{table*}[t]
	\centering
	\begin{tabular}{cc|c|c|c|c}
		\hline\hline
		&~~3PCDF~~ &~~5PCDF~ ~&~~3PADF~~&~~ 5PADF~~&~ ~Shooting\\
		\hline
		&$\kappa=-1~~~\kappa=1$&$\kappa=-1~~~\kappa=1$&$\kappa=-1~~~\kappa=1$&$\kappa=-1~~~\kappa=1$&$\kappa=-1~~~\kappa=1$\\
		\hline
		\small
		& $-55.006~~\boxed{-55.006}$   & $-55.005~~\boxed{-51.734}$&$-55.004~~-46.157 $&$-55.005~~-46.162 $&$-55.005~~-46.162 $\\
		& $\boxed{-46.165}~~-46.165$    &  $\boxed{-35.186}~~-46.162$&$-33.915~~-21.377$ &$-33.929~~-21.405 $&$-33.929~~-21.405$ \\
		&$-33.937~~\boxed{-33.937}$     &  $-33.930~~-21.405$&   $-9.171~~-0.284$ & $-9.210~~-0.304$  &$-9.210~~-0.259$ \\
		&$\boxed{-21.419}~~-21.419  $     &  $ -9.211~~~\boxed{-14.263}$&   $-~~~~~-$& $-~~~~~-$  &$-~~~~~-$  \\
		&$ -9.230~~\boxed{-9.230}$           &  $-~~~~~-0.290$&   $-~~~~~-$& $-~~~~~-$ &$-~~~~~-$  \\
		
		\hline\hline
	\end{tabular}
	\caption{ The neutron energy levels of $^{132}{\rm Sn}$ for $\kappa=-1$ and $\kappa=1$ in Wood-Saxon potential obtained by the FDM with different difference formulas and shooting method. The spurious states are marked in boxes. The unit of the energy level is MeV.}\label{table.1}
\end{table*}

In the following, we will apply the ADF to calculate the first-order derivative in the FDM to solve the Dirac equation.  Taking the 3-point formula as an example, the forward or backward difference formulas are
\beq  \label{eq:3-point-ADF}
&&\frac{df(r)}{dr}\simeq\frac{-3f(r)+4f(r+h)-f(r+2h)}{2h}, \nn 
&&\frac{df(r)}{dr}\simeq\frac{f(r-2h)-4f(r-h)+3f(r)}{2h}.
\eeq
One may notice that, if only forward or backward difference formula was used for both the large and small wavefunction components, the Dirac Hamiltonian matrix thus established would not be hermitian.  Actually, we found that for the wavefunction components with even parity, the backward difference formula should be applied to guarantee that its derivative is zero at $r=0$, according to the boundary condition. Instead, for the wavefunction components with odd parity, the forward difference formula should be used.   It should be also noticed that the parities of the large and small components of the same state in the Dirac equation are opposite.  Therefore, we should apply the forward or backward difference formula alternatively for the large and small components of the Dirac wavefunction according to their parities.  This prescription can not only guarantee the hermitian of the Dirac Hamiltonian, but also include the full wavefunction information while doing the first-order derivatives, and thus eliminate the spurious state.  

Explicitly, for the states with $\kappa=-1$, the large (small) component of the Dirac wavefunction should have 
the odd (even) parity.  Then, the up-right corner matrix $\bm B_1$ in the Dirac Hamiltonian (\ref{eq:Dirac-matrix-2}) which includes the first-order derivative of the small component should use the backward difference formula.  
Taking the 3-point formula (\ref{eq:3-point-ADF}) as an example, this matrix can be written as
\begin{widetext}
\begin{align}
\bm B_{1b}=
\begin{gathered}
	\begin{pmatrix}
		\frac{\kappa}{r_1}-\frac{3}{2h}&& & & & & & \\
		\frac{4}{2h}&\frac{\kappa}{r_2}-\frac{3}{2h}& & & & && \\
		-\frac{1}{2h}&\frac{4}{2h}&\frac{\kappa}{r_3}-\frac{3}{2h}& & & && \\
		& & & ...& & & &\\
		& & & &  -\frac{1}{2h}&\frac{4}{2h}&\frac{\kappa}{r_{n-1}}-\frac{3}{2h}&\\
		& & & &  &  	-\frac{1}{2h}&\frac{4}{2h}&\frac{\kappa}{r_n}-\frac{3}{2h}\\
	\end{pmatrix}
\end{gathered}
\end{align}
\end{widetext}
and the corresponding matrix $\bm B_{2}$ at the bottom-left of the Dirac Hamiltonian for the large component  should be calculated with the forward difference formula (denoted as $\bm B_{2f}$), which turns out to be the same with $\bm B^T_{1b}$. 


The first five bound states obtained by the FDM using the above 3-point ADF 
for the states with $\kappa=-1$ and $\kappa=1$ are listed in columns 3PADF in Table~\ref{table.1}.  
One can find that there is no spurious state anymore.  The results have one-to-one correspondence to those obtained by the shooting method.  If we use the 5-point ADF with higher accuracy, the results listed in columns 5PADF are much closer to those obtained by the shooting method.  The largest energy difference appears in the $3p_{1/2}$ state which is very weakly bound with the energy $E\approx-0.3$~MeV.  We have checked that if the box size is enlarged to be $R_{\rm box}=40$~fm, the 5PADF and shooting method will give the same results $-0.292$~MeV.   Table~\ref{table.1} shows that with a smaller box $R_{\rm box}=20$~fm, the FDM with 5PADF can give more accurate eigen energy for the weakly bound states than the shooting method.

\begin{figure}[htb]
	\centering
	\includegraphics[scale=0.38]{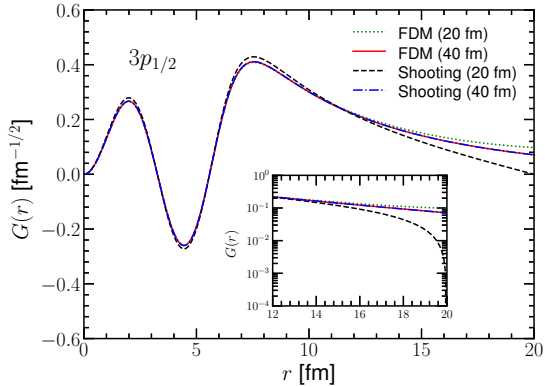}
	\caption{The wavefunctions of the large component $G(r)$ of $3p_{1/2}$ state obtained by FDM using 5PADF and shooting method with box sizes $R_\text{box}=20$ fm and $40$ fm. The inset shows the asymptotic wavefunction in the log scale.}\label{fig.2}
\end{figure}

 In Fig.~\ref{fig.2}, the large components of the wavefunction for $3p_{1/2}$ obtained by shooting method and FDM using 5PADF with box sizes $R_{\rm box}=20$ fm and $40$ fm are shown up to $r=20$ fm. When the box size is $R_{\rm box}=40$ fm, the wavefunctions obtained by the two methods are identical. With the box size $R_{\rm box}=20$ fm, the wavefunction obtained by the shooting method is obviously different from those obtained with the box size $R_{\rm box}=40$ fm especially in the asymptotic region.  Comparatively, the wavefunctions obtained by the FDM with the box $R_{\rm box}=20$ fm are much closer to those obtained with the box size $R_{\rm box}=40$ fm.  This shows that the FDM is less sensitive to the box size than the shooting method for the weakly bound state. {The reason is that the boundary condition for the wave function is $G(r)=0$ at $r=R_{\rm box}$ in the shooting method, but at $r>R_{\rm box}$ in the FDM}.
 
In summary, the spurious-state problem in the FDM to solve the Dirac equation for massive fermion is resolved directly without any evolution process or adding any other artificial terms.   The spurious states are completely eliminated by using the ADF instead of the CDF to calculate the first-order derivative in the Dirac Hamiltonian.  To guarantee the hermitian of the Dirac Hamiltonian, the forward and backward ADF should be used alternatively for the large and small components of the wavefunction according to their parities.  This prescription is illustrated by the example of neutrons moving in a Dirac Woods-Saxon potential in $^{132}{\rm Sn}$.  The feasibility of this prescription is also checked for the hydrogen system.  This prescription provides a very simple and efficient technique to apply the FDM for the description of the relativistic systems in the fields of atom physics, nuclear physics, particle physics, and so on. 

This work was supported in part by the National Natural Science Foundation of China (Grants  Nos. 11775119 and 2175109) and  the Natural Science Foundation of Tianjin (Grant  No. 19JCYBJC30800).

\end{document}